# High-resolution simulations of the final assembly of Earth-like planets 2: water delivery and planetary habitability

## Sean N. Raymond[1,2,4*], Thomas Quinn[2,4], and Jonathan I. Lunine[3,4]

1. NASA Postdoctoral Program Fellow, Center for Astrophysics and Space Astronomy, University of Colorado, Boulder, CO 80309-0389
2. Department of Astronomy, University of Washington, Box 351580, Seattle, WA 98195
3. Lunar and Planetary Laboratory, University of Arizona, Tucson, AZ 85287.
4. Member of NASA Astrobiology Institute
* Corresponding author: Email – raymond@lasp.colorado.edu, phone (303) 735 3729

## Abstract

The water content and habitability of terrestrial planets are determined during their final assembly, from perhaps a hundred 1000-km ``planetary embryos'' and a swarm of billions of 1-10 km ``planetesimals.''  During this process, we assume that water-rich material is accreted by terrestrial planets via impacts of water-rich bodies that originate in the outer asteroid region.  We present analysis of water delivery and planetary habitability in five high-resolution simulations containing about ten times more particles than in previous simulations (Raymond et al 2006a, Icarus, 183, 265-282). These simulations formed 15 terrestrial planets from 0.4 to 2.6 Earth masses, including five planets in the habitable zone.  Every planet from each simulation accreted at least the Earth's current water budget; most accreted several times that amount (assuming no impact depletion).  Each planet accreted at least five water-rich embryos and planetesimals from past 2.5 AU; most accreted 10-20 water-rich bodies.

We present a new model for water delivery to terrestrial planets in dynamically calm systems, with low-eccentricity or low-mass giant planets – such systems may be very common in the Galaxy.  We suggest that water is accreted in comparable amounts from a few planetary embryos in a "hit or miss" way and from millions of planetesimals in a statistically robust process.  Variations in water content are likely to be caused by fluctuations in the number of water-rich embryos accreted, as well as from systematic effects such as planetary mass and location, and giant planet properties.

# 1. Introduction

The last stages of terrestrial accretion consist of the agglomeration of a swarm of trillions of km-size planetesimals into a few massive planets (see the review by Chambers 2004 and references therein). This process determines the orbit, mass, and water content of terrestrial planets. These properties, in turn, determine the possibility that such planets may be hospitable for life.

What characteristics define a habitable planet? If the Earth is a good representation, then it appears that a habitable planet must satisfy the following criteria. 1) It must have a significant mass in order to maintain a relatively thick atmosphere. Long-term climate stabilization via the $CO_2$-carbonate cycle is induced by plate tectonics, which in turn requires a significant heat flux from radioactive nuclides. If this process is a requisite for life, then a very rough estimate of the minimum mass of a habitable planet is about 0.3 Earth masses ($M_\oplus$; Williams et al., 1997, Raymond et al., 2006b). 2) It must orbit its star in the circumstellar habitable zone, where the surface temperature is adequate to support liquid water on its surface (Kasting et al. 1993). 3) It must have an appreciable water budget – a dry planet in the habitable zone is not likely to be habitable. In addition, water may be an important factor for plate tectonics (Regenauer-Lieb et al., 2001).

Morbidelli et al. (2000) proposed that the bulk of Earth's water was accreted from a few water-rich embryos from the outer asteroid region, beyond 2-2.5 AU (a lesser amount of water, according to Morbidellie et al. (2000), was accreted as well from smaller bodies in the asteroid belt and comets). The water content of protoplanets[1] in the habitable zone is assumed to be very small, based on a number of considerations (Lunine, 2006; but see Drake and Righter, 2002), so an external source of water is needed. The work of Morbidelli et al. (2000), as well as earlier authors (e.g., Wetherill, 1996), showed that planetary accretion is a stochastic process, especially if much of the mass is contained in a small number of large embryos. The amount of water delivered to any given terrestrial planet, and its source region, can vary widely. If this view is valid, then it implies that terrestrial planets around other stars may differ greatly in terms of water amounts, and the timing of formation and location of giant planets from one system to another could play an important role in this variation (Lunine et al., 2003). Low resolution simulations conducted by Raymond et al. (2004) produce some terrestrial planets with little or no water and, hence, support this picture of potentially large variations in water abundance.

Previous dynamical simulations (Wetherill, 1996; Morbidelli et al., 2000; Chambers and Cassen, 2002; Levison and Agnor, 2003; Raymond et al., 2004, 2005a, 2005b; Raymond, 2006), though they start from only 20-200 particles, established the following trends relating to planetary habitability. 1) The eccentricity of the giant planets has a strong effect on the water content of the terrestrial planets (Chambers and Cassen, 2002; Raymond et al., 2004; Raymond, 2006). 2) An increase in the giant planet mass results in the formation of a smaller number of more massive terrestrial planets (Raymond et al., 2005a). With fewer planets forming, the chances that a given planet will fall in the habitable zone is reduced. 3) A disk with an increased surface density of solid material will form a smaller number of more massive terrestrial planets with larger water contents than lower-mass disks, due to stronger self scattering among protoplanets (Raymond et al., 2004, Raymond et al., 2006b).

---

[1] We use the term protoplanets to refer to all terrestrial building blocks, including both planetesimals and planetary embryos. This departs from certain previous uses of the term, which use protoplanets as a synonym for planetary embryos.

In this paper, we focus on water delivery and planetary habitability in five high-resolution simulations from Raymond et al. (2006a) that contain between 1000 and 2000 initial particles. For the first time, these simulations directly simulate a realistic number of embryos according to various models of their formation. The simulations reported here were designed to examine the accretion and water delivery processes in more detail, and also to explore the dynamical effects of including a larger number of particles than in previous simulations.

Several researchers have shown that the accretion process is stochastic; two simulations with almost identical initial conditions can form planetary systems that are quite different (e.g., Chambers 2001). In choosing initial conditions, we have constructed systems that sample a few different models. Due to computational constraints, however, we performed only one simulation of each case. Thus, we were not able to examine the stochastic variations between similar simulations or explore parameter space. Our five simulations, however, represent an ensemble of high-resolution simulations by which we consider, in detail, water delivery in a limited scenario.

Section 2 briefly summarizes the simulations from Raymond et al. (2006a). Section 3 discusses the delivery of water-rich material to the terrestrial planets and includes a discussion of the statistical robustness of the process. Section 4 explores the physical properties and potential habitability of the planets formed in each simulation. Section 5 summarizes our new results and concludes the paper.

**2. Description of simulations**

Our dataset consists of the five simulations from Raymond et al. (2006a), which were started from three sets of initial conditions. Each system started from a disk of 1000-2000 protoplanets totaling 8.6 to 9.9 $M_\oplus$. Protoplanets were placed on low-eccentricity and low-inclination orbits between 0.5 and 5 AU, following a surface density distribution that varies with radial distance $r$ as $r^{-3/2}$. Each simulation included a Jupiter-mass planet on a circular orbit at 5.2 or 5.5 AU (hereafter simply referred to as ``Jupiter''). Note that our starting number of particles was ~5-10 times larger than in previous work.

- Simulation 0 started in the late stages of oligarchic growth, when planetary embryos were not yet fully formed. It contained a total of 1885 bodies, with masses from between roughly $10^{-3}$ and $10^{-2}$ $M_\oplus$. Jupiter was included at 5.5 AU, outside Jupiter's current position to account for the giant planet's inward drift as it ejected rocky protoplanets (so that it might end up at 5.2 AU). However, in each simulation, the inward drift was only 0.05-0.1 AU.
- Simulation 1 contained 36 Moon- to Mars-mass planetary embryos out to 2.5 AU (the location of the 3:1 mean motion resonance with Jupiter at 5.2 AU), and 1000 planetesimals of 0.006 $M_\oplus$ each from 2.5 to 5 AU. This follows from the results of Kokubo and Ida (2000, 2002), who suggest that the timescale for the formation of planetary embryos is a function of heliocentric distance. In their models, the timescale for embryo formation at 2.5 AU is several millions of years, which is comparable to the timescale for the dissipation of the gaseous component of the Solar Nebula (Haisch et al., 2001, Briceno et al., 2001). Since giant planets are constrained to form in the presence of nebular gas, Kokubo and Ida's model suggests that embryos may not have fully formed past 2.5 AU before Jupiter's formation.

- Simulation 2 assumed that planetary embryos formed all the way out to 5 AU by the time of giant planet formation, following from alternate models of oligarchic growth that predict faster embryo growth (Weidenschilling et al., 1997, Goldreich et al., 2004). Thus, simulation 2 contained 54 Moon- to Mars-mass planetary embryos out to 5 AU. A significant component (roughly one third) of the total mass was contained in a swarm of 1000 planetesimals of 0.003 $M_\oplus$ each that also extended from 0.5 to 5 AU.

For simulations 1 and 2, we performed two runs in which we varied the numerical treatment of planetesimals. In simulations 1a and 2a, all bodies interacted with each other. In simulations 1b and 2b, the ``planetesimals'' were not self-interacting. They gravitationally interacted with embryos and with Jupiter, but not with each other, which allowed for significant computational speedup. Each simulation was evolved for ≥200 Myr with the hybrid integrator *Mercury* (Chambers 1999) using a timestep of 6 days.

Our sets of starting conditions reflect the wide range of plausible values for the formation timescales of both giant planets and planetary embryos. Simulation 1 reflects a scenario in which Jupiter is fully formed before embryos have formed in the outer asteroid region (2.5 to 5 AU). This corresponds to either relatively fast giant planet formation (e.g., Alibert et al., 2004) or relatively slow embryo formation (Kokubo & Ida 2000, 2002, Leinhardt & Richardson 2005).[2] In contrast, the initial conditions of simulation 2 inherently assume that embryos in the outer asteroid region form more quickly than giant planets (e.g., Weidenschilling et al., 1997, Goldreich et al., 2004). Simulation 0 started from the later phases of oligarchic growth, before any embryos have reached their isolation mass, and therefore assumes that the timescales for giant planet and embryo formation are comparable.

Our choice of a circular Jupiter allows us to study radial mixing and water delivery induced by interactions between protoplanets with relatively weak giant planet perturbations. Indeed, it has been shown that an eccentric Jupiter preferentially ejects much of the water-rich material beyond 2.5 AU, which causes the terrestrial planets to be dry (Chambers and Cassen 2002, Raymond et al., 2004). It has also been shown that, for water-rich terrestrial planets to form in the habitable zone, a Jupiter-mass giant planet must be at least 3.5 AU from the star, and much farther if its eccentricity is nonzero (Raymond 2006). A Jupiter-mass giant planet at 5 AU, even on a circular orbit, plays a negative role in the water delivery process, ejecting more water-rich material than it scatters inward (Raymond et al., 2005a). Our initial conditions also reflect a new model for the early dynamical evolution of the outer planets in the Solar system, in which the giant planets' eccentricities were negligible during the time of terrestrial planet formation (Tsiganis et al., 2005). In this model, Saturn's influence on the terrestrial planets is negligible, because secular resonances such as the ν6 at 2.1 AU did not yet exist. Thus, our simulations are well-suited for studying the process of water delivery under the assumption of relatively small influence from the giant planets. This may be relevant for our Solar System, but not to the subset of extra-solar planetary systems with very eccentric orbits, or those with close-in giant planets (see, Fogg and Nelson, 2005, Raymond et al., 2006c). However, the abundance of environments such as the one studied here is uncertain. Given that our simulations contain only one giant planet on a circular orbit, we consider these systems to be "dynamically calm." Related, calm systems include lower-mass (e.g., Neptune-like) giant planets whose perturbations are weaker and could, therefore, have nonzero eccentricities. The population of extra-solar giant planets has a median eccentricity of about 0.2 (Butler et al., 2006). However, systems with only Neptune-mass planets are now being discovered (Lovis et al., 2006). In addition, the presence of debris disks does not correlate with known giant planets (Greaves et al., 2006). Thus, many terrestrial planets may be forming in environments with no large perturber – such environments may also be considered dynamically calm. Note that several strong resonances and giant planet secular forcing still pervade our simulations. So, we cannot directly apply our simulations to systems with no giant planets.

---

[2] Slow embryo growth is not inconsistent with the "core-accretion" model for giant planet formation (e.g., Pollack et al 1996). If there exists a significant density increase past the snow line, then embryos (and giant planet cores) could form more quickly in the giant planet zone than in the outer asteroid region's lower density.

The initial water content of protoplanets in our simulations was designed to reproduce the water content of chondritic classes of meteorites (Abe et al., 2000; see Fig. 2 from Raymond et al., 2004): inside 2 AU, bodies were initially dry; outside 2.5 AU, they had an initial water content of 5% by mass; and between 2 and 2.5 AU, they contained 0.1% water by mass. The starting iron contents of protoplanets were interpolated between the values for the planets (neglecting the planet Mercury) and chondritic classes of meteorites, with values taken from Lodders and Fegley (1998), as in Raymond et al. (2005a, 2005b). To span our range of initial conditions, we interpolated to values of 0.5 at 0.2 AU and 0.15 at 5 AU.

Each simulation took between four and sixteen months to run on a fast Linux PC. Fifteen terrestrial planets formed in these five simulations, including five potentially habitable planets, whose properties are listed in Table 1. The typical eccentricities of planets were about 0.05, which was lower than in previous simulations but still somewhat larger than for the Solar System's terrestrial planets (see Raymond et al. (2006a) for a discussion). The high-eccentricity planets that formed in simulation 2b were a surprise given the high resolution of these simulations. However, as shown in Fig. 20 of Raymond et al. (2006a), the number of planetesimals dwindled to zero after about 50 Myr, so dynamical friction was very weak in the final stages of formation. We do not expect higher resolution simulations to form such high-eccentricity planets.

## 3. Water Delivery to Terrestrial Planets

Water is thought to be an essential ingredient for life. In addition, water lowers the viscosity of rocks and may be an important factor in determining whether a planet will develop plate tectonics (Regenauer-Lieb et al., 2001). Plate tectonics, in turn, plays an essential role in climate stabilization through the carbon cycle (Walker et al., 1981). The large water contents of the planets (listed in Table 1) were not unexpected, since our initial conditions contained about 50% more mass than the minimum mass solar nebula (e.g., Hayashi 1981) and only one giant planet on a circular orbit.

The water content of the Earth is uncertain to within a factor of a few. The mantle contains between 1 and 10 oceans of water (see Lecuyer, 1998, or Morbidelli et al., 2000, for a discussion), where an ocean is defined as the amount of surface water on the Earth, $1.5 \times 10^{24}$ grams, or roughly $2.5 \times 10^{-4}$ Earth masses. In earlier papers (Raymond et al., 2004, 2005a, 2005b) and in Tables 1 and 2, we assumed the Earth's total current water budget to be four oceans (one ocean on the surface and three in the mantle), which corresponds to a water mass fraction of $10^{-3}$.

Here, we describe in detail the acquisition of water by terrestrial planets in our simulations. We assume that protoplanets that form at 1 AU are dry and the water content of chondritic classes of meteorites is representative of the starting water content of protoplanetary disks. In section 3.1, we summarize the sources and timescales for water delivery to terrestrial planets, including comets. In section 3.2, we discuss the efficiency of the delivery of water-rich material from the outer asteroid region. In section 3.3, we discuss the mass distribution of water-bearing impactors, with consequences for the statistics of the water delivery process. In section 3.4, we discuss the retention of water during collisions.

*3.1 Summary of sources of water and delivery timescales*

In the simulations presented here, we only considered the outer asteroid region as a source of water. However, the Earth likely accreted water-rich material from several sources at different times (see Conclusions from Morbidelli et al., 2000 for a more comprehensive discussion):

1. Icy planetesimals in the Jupiter-Saturn region have dynamical lifetimes of only a few hundred thousand years and Earth-collision probabilities of only ~$10^{-6}$ (Morbidelli et al. 2000), so these impacted the terrestrial planets early in their formation, very soon after giant planet formation. If Jupiter formed quickly, most of this water was likely lost due to the planets' small masses (and therefore low surface gravities). At most, icy bodies at and beyond the Jupiter-Saturn region could have contributed about 10% of Earth's water (Morbidelli et al. 2000).

2. During accretion, water-rich planetesimals and embryos from the outer asteroid region between 2.5-4 AU were incorporated into the planets, delivering the bulk of Earth's water. This is the source of water considered in our simulations.

3. After accretion, icy comets from beyond Neptune's orbit were scattered into the inner Solar System. The ``late heavy bombardment'' of asteroids and comets occurred roughly 700 million years after the terrestrial planets formed (e.g. Gomes et al. 2005). A comparison of the D/H ratio in Earth water and comets suggests that comets contributed, at most, 10% to the Earth's water budget[3], which is consistent with the dynamical results (Morbidelli et al. 2000).

4. The Earth continues to be impacted at a low rate by water-rich asteroids and comets (e.g., Levison et al., 2000).

*3.2 Source regions and efficiency of water delivery*

Figure 1 (top panel) shows the fraction of water-rich material delivered to the surviving terrestrial planets as a function of starting semimajor axis. The region from 2-5 AU is divided into six bins, each with a width of 0.5 AU (i.e., 2-2.5 AU, 2.5-3 AU, etc). We only consider water delivery to planets inside 2 AU, thereby excluding planet *d* from simulation 2a. It is clearly easiest to deliver water to the terrestrial planets from the innermost water-rich region, between 2 and 2.5 AU. The efficiency of water delivery drops off at higher orbital distances, because (i) bodies are physically more distant and need to travel greater distances to impact the terrestrial planets, and (ii) Jupiter's dynamical effects cause many asteroidal bodies to be ejected before they may be accreted onto the terrestrial planets (e.g., Chambers and Cassen 2002). The efficiency of water delivery was similar for most simulations, with typical mid-asteroid belt region values between 5% and 20%. The efficiency of water delivery in simulation 2b was very high, with values above 20% out to 4 AU, but recall that this simulation was anomalous in the high eccentricities of the planets that formed. Simulation 2a had the lowest values, because water delivery to planet *d* at 2.19 AU was not included. Planet *d* acted as a small dynamical barrier for inward-diffusing bodies, and also accreted much of its local, water-rich material.

---

[3] Note that the noble gas ratios in the Earth's atmosphere are not consistent with those in asteroidal material; however, this is explained by a dual source of Earth's atmosphere -- a mixture of nebular and chondritic components (Dauphas 2003, Genda and Abe 2005).

Figure 1 (bottom panel) shows the total amount of water delivered to the terrestrial planets from each bin between 2 and 5 AU for each simulation. Again, this only includes planets inside 2 AU, which excludes planet *d* from simulation 2a. The shape of the curves mirror those in the top panel, except for the innermost bin that represents water delivery from 2-2.5 AU. The water content of this region is much lower than past 2.5 AU (0.1% vs. 5%), so the region simply does not contain much water. The total starting water content between 2-2.5 AU was about 3 oceans in each simulation. The major source of water for terrestrial planet lies between 2.5 and 4 AU. Inside 2.5 AU, bodies don't start with enough water, and past 4 AU the dynamical lifetime of bodies is too short to deliver water. Indeed, the vast majority of material exterior to the 3:2 mean-motion resonance with Jupiter is destroyed via collisions with or ejections by Jupiter in the first $10^{4-5}$ years of each simulation. Note that, if Jupiter and Saturn were on their current orbits during terrestrial planet formation, the shape of the curves in Fig. 1 may look different because of the presence of the $\nu_6$ resonance at 2.1 AU (see, e.g., O'Brien et al., 2006). In addition, even the giant planets' current modest eccentricities may affect water delivery (Chambers and Cassen 2002; Raymond et al., 2004).

Figure 2 shows the water mass fraction of each terrestrial planet in our five simulations as a function of the planet's final mass. Higher-mass planets do not necessarily have higher water contents than lower-mass ones. Indeed, Fig. 2 is more or less a scatter plot. Almost all planets have water mass fractions between $3\times10^{-3}$ and $2\times10^{-2}$. Figure 3 shows that the water mass fraction is a function of the planets' orbital distance, displaying a slight trend toward higher water contents at larger orbital distances. As shown in Raymond et al. (2004), planets with orbital radii inside 1 AU accrete systematically less water. The water content of the planets formed in these simulations was a much stronger function of orbital distance than of planetary mass, though there was significant scatter.

*3.3 Statistics of water delivery*

The size distribution of water-bearing impactors is important for the statistical robustness of the water delivery process, as well as for volatile retention during collisions. We have the resolution in these simulations to discriminate between planetary embryos and planetesimals, and to make a rough estimate of the number of bodies involved in the water delivery process. Note, however, that our planetesimals are still many orders of magnitude larger than realistic values. In addition, we have only one simulation for each starting condition, so we can not see stochastic variations between similar simulations. Nonetheless, our simulations are the best to date and warrant a detailed examination.

Figure 4 shows the timing of all impacts that built up the three planets from simulation 0, color-coded by planet. In time, the feeding zones of each planet both widened and moved outward, eventually reaching water-rich areas. Note that the "accretion seeds" of planets *a* and *b* – in some sense the core embryos of each planet[4] – originated in very similar locations, at 0.8825 AU (planet

---

[4] To be more specific, the accretion seed can be defined as follows (see also Morbidelli et al 2000). When two objects collide in the simulation, the remaining agglomeration keeps the name of the larger impactor while the name of the

*a*) and 0.8660 AU (planet *b*). Thus, the feeding zones of these two planets overlap at early times, most notably in the first 5 x $10^5$ years of the simulation.

In Fig. 4, it can be seen that water-rich bodies were not accreted by the planets in this simulation until about 10 Myr after the start of the simulation. We believe the reason for this delay has to do with a change in the mass spectrum of bodies as they accreted and grew; until bodies of a certain size exist in a given region, gravitational scattering events are too weak to alter significantly a protoplanet's orbit. In addition, a large amount of small bodies remains to damp eccentricities and keep adjacent feeding zones separate. In time, the mass of protoplanets grows and their numbers decrease, causing stronger gravitational scattering and less dynamical friction. Thus, feeding zones are widened and eventually overlap, resulting in radial mixing in the disk and the delivery of water-rich material from the outer asteroid region to the growing terrestrial planets (see Raymond et al. (2006a) for more discussion).

Each planet in our five high-resolution simulations accreted at least 5 water-rich bodies and was delivered at least 6 oceans of water; most accreted more than 10 water-rich bodies and up to 50-100 oceans. Table 2 shows the water content of each planet, as well as the number of water-rich impactors and the fraction of water delivered in the form of large bodies (embryos) vs. small bodies (planetesimals). Since our planetesimals are much larger than realistic, 1 km planetesimals, we must be careful in our definition of ``small bodies.'' This distinction is only needed for simulations 0 and 1a, as in the other cases water-rich embryos were either included at the start of the simulation (simulations 2a, 2b) or could not themselves accrete (simulation 1b). The smallest planetary embryos that formed, given a typical spacing of 5-10 mutual Hill radii, were 0.03-0.05 $M_\oplus$ in the very inner part of the disk. Embryos past 2 AU ranged from 0.07 to almost 0.2 $M_\oplus$, depending on the spacing. For simulations 0 and 1a, we therefore define a planetary embryo to be a body that is at least 0.05 $M_\oplus$ or an agglomeration of at least four smaller bodies. Note that there still exists a continuum in the mass range of water-bearing bodies, ranging from planetesimals that had not accreted any other bodies to Moon-size accumulations of 2-3 planetesimals.

Water-rich embryos are preferentially accreted by more massive planets and by planets at larger orbital radii. The path of a water-rich body from past the snow line to the terrestrial region involves an inward diffusion via multiple gravitational encounters with other bodies. The probability of such a body being accreted depends on the body's cross section for impact as well as the accreting body's cross section (i.e., its mass). The cross section of planetesimals is small, so the impact probability depends only on the cross section and orbit of the accreting body. More massive planets naturally accrete more water-rich material than less massive ones. Planets at larger orbital radii (at 1-2 AU rather than inside 1 AU) are simply closer to the source of water, and have a higher number of encounters than planets closer to the Sun. Water-rich embryos have a higher cross section for interaction than water-rich planetesimals, simply because of their larger masses. Thus, statistically speaking, we expect that water-rich planetesimals should be able to "diffuse" in to smaller orbital distances than water-rich embryos because of their smaller cross sections. Although our sample size is small, this is indeed what we observed in terms of the water budgets of our simulated planets (Table 2).

Figure 4 shows the amount of water delivered to the terrestrial planets in each impact, color-coded by simulation. The size of each body represents the relative physical size of each impactor. The

---

smaller impactor is lost. The planets that remain at the end of the simulation thus bear the name of the object which was the larger in its first collision, as well as every subsequent collision. This particle is called the "accretion seed".

amount of water per planetesimal in simulations 1a and 1b, ~1.2 oceans, is evident as a horizontal line that represents many accretion events. There clearly exists a range in the volume of water delivered during these impacts. The size of the impactor does not always correlate with the amount of water delivered. In some cases, large embryos either accreted one or two water-rich planetesimals, or perhaps formed in the ``slush'' region between 2 and 2.5 AU. These large bodies delivered only a very small amount of water. In general, however, larger impactors delivered more water, as they usually contained many water-rich planetesimals or originated past 2.5 AU themselves.

Fig. 4 shows that water-delivering impacts did not occur in simulation 0 until at least 10 Myr, at which point planets were likely to have reached a substantial fraction of their final mass and have primitive atmospheres. However, Fig. 5 shows that water-bearing impacts happened earlier in the other simulations. As discussed above, a given annulus remains dynamically isolated until it can grow bodies of a significant size, capable of scattering material out of the region. Simulation 0 was the only case in which we allowed this process to occur throughout the disk, and it displayed the expected, outward-moving trend. We, therefore, consider the trend in Fig. 4 to be our most accurate representation of accretion through time; indeed, simulation 0 is by far the most intensive computation of its kind that has been run to date. However, given that we did not include the effects of very small bodies (i.e., planetesimals of realistic size), it is possible that the timescale from Fig. 4 could be too long. In other words, we expect a long delay between the start of accretion and the onset of significant radial mixing. However, given the limited resolution in even our best simulation, we cannot accurately resolve the early formation of embryos.

The timing of water-bearing impacts within our simulations has important applications for comparison with the Earth's water budget. During core formation, siderophile elements follow iron into the core. Thus, the extent to which the abundance of siderophiles in the mantle constrains the amount of exogenous material that impacted the Earth depends on whether the material is accreted before or after core formation (Drake and Righter 2002; Nimmo and Agnor, 2006). Table 2 lists the amount of material that originated past 2.5 AU and was accreted after the last giant impact by the planets in each of our simulations. In most cases, this carbonaceous veneer contributed, at most, a few percent of a planet's total mass, which is consistent with the results of Morbidelli et al. (2000) and somewhat above the 1% post-core-formation specified by Drake and Righter (2002). This behavior was also seen in the "CJS" simulations of O'Brien et al.,(2006), which were similar to our simulation 2b. Interestingly, O'Brien et al.,(2006) found that far too much carbonaceous material accreted after core formation in simulations with an eccentric Jupiter and Saturn. Although uncertainties remain, this may lend credence to the model of Tsiganis et al.,(2005), in which the giant planets' orbital excitations were very small during terrestrial planet formation.

It is interesting to note from Table 2 that the size distribution of water-bearing impactors differs from planet to planet, even sometimes among planets in the same simulation. For example, in simulation 0 planet *a* accreted more than 90% of its water in the form of small bodies, while water-rich embryos contributed roughly two thirds of planets *b*'s and *c*'s water. Planets in the habitable zone and those farther out received a larger amount of water from embryos. However, the water contribution from planetesimals was significant in almost all cases; even in simulation 2b, in which the total mass in embryos was twice the total planetesimal mass, 2-3 oceans of water were delivered to each planet in the form of planetesimals.

In our simulations, small bodies played an important role in water delivery and contributed an amount of water roughly comparable to that from embryos. Both embryos and planetesimals tended to contribute roughly $4 \times 10^{-3}$ Earth masses (~15 oceans) of water to Earth-like planets, though there was clearly a lot of scatter. In contrast, Earth analogs in Morbidelli et al.,(2000) accreted an average of 0.1-0.2 Earth masses of water from embryos and only $5 \times 10^{-4}$ $M_\oplus$ in planetesimals (called "primitive asteroids" in that paper). The difference in water accretion from embryos between our simulations and those of Morbidelli et al.,(2000) are not very large and probably due to a combination of dynamical friction, the assumed initial water content of asteroidal material (we assume 5% beyond 2.5 AU whereas Morbidelli et al. assume 10%), and our initial conditions, which start with only about half to two thirds of Morbidelli et al.'s total mass in embryos. We are not certain of the origin of the order of magnitude discrepancy between the water delivery from our planetesimals and Morbidelli et al's. We suspect that it is due to differences in the dynamical treatment of small bodies. Our simulations included interactions between embryos and planetesimals, which resulted in feedback between the dynamical friction felt by embryos due to planetesimals and the orbits and impact rates of those planetesimals. In contrast, Morbidelli et al. treated planetesimals as massless particles, integrated under the influence of embryos and the giant planets, with no dynamical friction. With no dynamical friction and, therefore, higher embryo eccentricities, the excitation of planetesimals would be stronger than in our simulations, the planetesimal eccentricities and inclinations would be higher, and the mean planetesimal lifetime would be shorter, thereby reducing the collision probability with the terrestrial planets.

Our simulations incorporated realistic distributions of embryos, in terms of their number and masses. However, the masses of our planetesimals were roughly $5 \times 10^{-3}$ $M_\oplus$, about 7-9 orders of magnitude more massive than real 1-10 km planetesimals. What would be the effects of a realistic swarm of trillions of planetesimals? We are confident that our calculation of dynamical friction is correct, given the large ratio of the embryo or planet mass to the planetesimal mass, which was typically 20-100, depending on the simulation and the orbital zone (our embryo-planetesimal mass ratio was comparable to the simulations of O'Brien et al. (2006)). We expect that, in a true swarm of planetesimals, dynamical friction would last longer – by the end of our simulations the number of particles dwindled to less than 10 (in contrast to the million or so asteroids larger than 1 km). Note, however, that high-velocity collisions among km-size bodies may be disruptive and create a trail of impact debris – the effect of this debris may actually enhance dynamical friction (e.g., Goldreich et al., 2004). In addition, the number of water-bearing impactors would increase drastically for a true planetesimal swarm. In our simulations, the planetesimal distribution was dominated by the effects of embryos and the giant planet. So, we expect that the characteristic evolution of real, km-size bodies would follow the evolution we observed, in terms of the statistical orbital evolution of the swarm and the fraction of planetesimals entering the inner solar system. So, we expect the number of water-bearing planetesimals to scale with the number of simulated planetesimals.

We propose the following picture of water delivery: terrestrial planets accrete water from two reservoirs – planetary embryos and planetesimals. A comparable amount of water is accreted from a few embryos and many millions of planetesimals. In environments similar to the dynamically calm one studied here, we expect that small-number statistical variations will lead to fluctuations in the amount of water accreted by embryos. However, given the vast number of planetesimals in a real disk, we expect that the accretion of these bodies will not be dominated by statistical fluctuations. Thus, we propose that the total amount of water accreted by a given planet should vary based on the number of embryos accreted, while the planetesimal contribution should remain roughly constant, modulo certain systematic effects.

During the final stages of growth (after ~10 Myr), we envision the terrestrial planet zone as containing a diffuse swarm of water-rich planetesimals, in addition to growing embryos and a larger number of local, dry planetesimals.  The source of the icy planetesimal swarm is the region from 2-4 AU (Fig. 1).  The evolution of each body in this swarm is such that water-rich planetesimals undergo a random walk in orbital distance due to interactions with embryos until they either collide with a growing planet or leave the terrestrial planet region (back to the asteroid region or into the Sun).  As described above, collisions with more massive planets are more likely because of their larger physical size (there is a negligible amount of gravitational focusing because of the high relative velocities of water-rich planetesimals).  In addition, the water-rich swarm is less dense at smaller orbital distances because 1) the number of interactions needed to enter a given zone on a moderately eccentric orbit increases for smaller orbital distances, and 2) highly eccentric orbits (due to strong encounters) that enter the inner terrestrial zone spend only a small fraction of their time in that inner zone.  The planets in our simulations showed little to no trend in terms of their water content as a function of mass (Fig. 2).  However, Raymond et al., 2004 showed a statistically significant drop in the water contents of planets inside 1 AU, and we see a similar trend in Fig. 3.  Therefore, we believe that the water contents of planets are a much stronger function of orbital distance than of planet mass, at least in the ranges explored here.

In environments that are not dynamically calm (e.g., with eccentric giant planets or following giant planet migration), the details of water delivery may be significantly different.  Indeed, Chambers and Cassen (2002) and Raymond et al., 2004 showed that an eccentric Jupiter preferentially ejects water-rich material and causes the terrestrial planets to be dry.  O'Brien et al. (2006) confirmed this with newer, high-resolution simulations.  Raymond (2006) established limits on the orbital distance and eccentricity of a giant planet for water delivery to occur – for a circular orbit, a Jupiter-mass planet at 3.5 AU allows water-rich terrestrial planets to form in the habitable zone.  However, for an eccentricity of 0.1, the limit is 4.5-5 AU (Raymond 2006).

The fraction of water accreted in the form of planetesimals vs. embryos varies from simulation to simulation and reflects the same stochastic nature of embryo accretion pointed out by the simulations of Morbidelli et al. (2000). However, because the number of particles they could incorporate was limited to a few hundred, the mass range of planetesimals was restricted to lunar- to Mars-size.  The planets in simulations 2a and 2b accreted a much larger amount of water as embryos than in our 3 other simulations.  Indeed, each planet but one (planet *a* from simulation 2a) received >90% of its water from embryos (though each planet accreted at least 1 ocean from planetesimals, and typically a few).  An important aspect of our proposed model is the amount of water delivered by planetesimals.  Do small bodies deliver 5-10% of the total water (as in simulations 2a and 2b) or closer to 50% (as in simulations 0, 1a and 1b)?  We began our simulations with roughly two thirds of the mass in the form of embryos, though oligarchic growth should end when half of the mass is in large bodies.  A larger fraction of planetesimals should increase the relative amount of water-rich small bodies involved in water delivery.  In addition, embryos are thought to grow more slowly at larger heliocentric distances (e.g., Wetherill and Stewart 1993, Kokubo and Ida 2000), though a density increase beyond the "snow line" is thought to shorten the embryo formation time in the giant planet region – this is vital for the core-accretion model of giant planet formation (e.g., Pollack et al., 1996).  Since dynamical and collisional timescales are shorter in the outer asteroid region (2.5-4 AU) than in the terrestrial planet region (0.5-1.5 AU), it is reasonable to expect a larger fraction of the total asteroidal mass to be in the form of planetesimals than at 1 AU. So, although we cannot quantify it with the present simulations, we expect that the water contribution from small bodies could be comparable to that from the embryos, but strongly dependent on the initial conditions.

*3.4 Retention of water during accretion*

If a planet has an atmosphere, then the explosion generated by the impact of a planetesimal-size body is confined and very little material escapes (Sleep and Zahnle 1998). In the absence of an atmosphere, the volatile retention of the impact of a planetesimal-size body is largely determined by the impact velocity and the planet's escape velocity (Sleep and Zahnle, 1998; Segura et al., 2002). Late stage, water-rich planetesimal impacts have typical impact velocities of a few to about 30 *km s$^{-1}$* more than the escape speed of the accreting body (Fig. 6 from O'Brien et al., 2006). In this velocity range, we expect the bulk of impacting bodies and their volatiles to be preserved on the accreting planets, though this depends somewhat on the planet's escape speed (Sleep and Zahnle 1998). Thus, volatiles from water-bearing planetesimal impacts at early times when the planets are relatively small may be only partially retained. However, at later times the bulk of water from small bodies should be retained.

A collision between an Earth-size planet and a Mars-size body will typically erode 30% of the larger planet's atmosphere and 10% of the smaller planet's atmosphere (Genda and Abe 2003). A key factor in the retention of the planet's atmosphere and ocean in a giant impact is the ground velocity induced by the impact shock. Genda and Abe (2005) showed that the presence of an ocean greatly affects a planet's ability to retain an atmosphere or the oceans themselves. The shock impedance of an ocean is less than that for a planetary surface, so the ground velocity during an impact on an ocean-covered world is higher, as is the escape fraction. The highly energetic vapor injected from the oceans into the atmosphere during an impact imparts energy to the atmosphere, which may exceed the escape velocity and leave the planet (Genda and Abe 2005; Zahnle 2005). Only when the atmosphere is destroyed can water be lost from the planet's surface (Genda and Abe 2005).

The time dependence of water-delivering impactors seen in Fig. 4 implies that those planets' water retention may have been a function of time. We expect this to be true in general for accreting terrestrial planets, for at least two reasons: 1) the planets' escape velocities increase as they grow; 2) water-bearing impacts tend to preferentially occur late in the accretion process. The earliest water-delivering bodies encounter dry planets with strong impedance to the impact-generated shock wave, so water and the atmosphere should be mostly retained. Later water-bearing bodies may impact ocean-covered worlds, and could potentially erode the planet's atmosphere and oceans (Genda and Abe, 2005).

Additional planetary water is thought to be lost after accretion. Matsui and Abe (1986) and Kasting (1998) showed that if Earth and Venus had the same starting water content, Venus would lose most of its water post-accretion via hydrodynamic escape during a post-collision, hot water vapor atmosphere phase. Thus, a planet's final water content is a complex combination of the nature of the accretionary impacts that built it and the physical characteristics of its location (in particular, the temperature).

## 4. Physical properties and planetary habitability

Our five simulations formed a total of fifteen terrestrial planets more massive than 0.3 $M_\oplus$; fourteen of these had semimajor axes inside 2 AU, and five lay in a broadly-defined habitable zone of 0.9-1.4 AU (see discussion below)[5]. Table 1 lists rough physical properties of all surviving planets, including an estimate of the planetary radius, assuming that radius scales as mass$^{0.27}$ (Valencia et al., 2006). Figure 6 shows the final configuration of all five simulations, with the Solar System included for scale. All have water contents equal to or higher than the Earth's current water content, without taking volatile loss into account. Our simulations are only a very rough representation of a very complex process that involves additional physics and trillions of bodies. Nonetheless, our artificial planets give a glimpse of the possible diversity of terrestrial exoplanets. If such planets were discovered around other stars, would any of them be suitable for life?

The mean eccentricities listed in Table 1 are averaged over many millions of years and do not capture fluctuations on shorter timescales. For most planets, eccentricity variations are significant but not exceedingly large. For example, on Myr timescales the Earth's eccentricity oscillates between about 0.0 and 0.06 (Quinn et al., 1991); planet 0-*b*'s eccentricity ranged from 0.005 to 0.095. In terms of planetary habitability, we expect that eccentricity variations are important for planets that lie on the edge of the habitable zone, experience rather large amplitude variations, and thus may spend a portion of their orbits outside the habitable zone (e.g., Williams and Pollard 2002).

This was the situation for the two potentially habitable planets of simulation 2a. Planet 2a-*b* lay at the inner edge of the habitable zone at 0.94 AU, and planet *c* skirted the outer edge of the habitable zone at 1.39 AU. The eccentricities of both planets oscillated with amplitudes of 0.1-0.2 on a timescale of a couple hundred thousand years and were dynamically linked; when planet 2a-*b* was in a high-eccentricity state, planet *c* was in a low-eccentricity state, and vice versa. Figure 7 shows the eccentricity evolution of the two planets as a function of time for a 2 Myr interval late in the simulation. The amplitude of planet 2a-*b*'s eccentricity variation was larger than planet 2a-*c*'s, and the variations of each planet were exactly out of phase. Planets 2a-*b* and 2a-*c* were in *apsidal libration* -- over long timescales the orientation of their orbits librated about anti-alignment with an amplitude of about 60°. This kind of dynamical behavior is common in systems of two or more planets (e.g., Gladman 1993).

Figure 8 shows the orbits of planets *b* (in red) and *c* (in blue) from simulation 2a at two different times. During time 1 (left panel), planet *b*'s eccentricity was low and planet *c*'s was high. During time 2 (right panel), planet *b*'s eccentricity was high and planet *c*'s was low. The two orbital configurations shown in Fig. 8 show the most extreme eccentricity configurations shows the most extreme eccentricities values found for the planets over the last 50 Myr of the simulation (see Fig. 7). The anti-alignment of longitudes of pericenter is evident as planet *b*'s perihelion at time 2 was on the opposite side of the Sun from planet *c*'s perihelion at time 1.

At time 1, planet *b*'s orbit stayed at the inner edge of the habitable zone, but planets *c* strayed far past the outer edge, to its aphelion of 1.61 AU. At time 2, planet *b* ventured far inside the inner edge of the habitable zone, with a perihelion distance of 0.73 AU, while planet *c* skirted the outer edge of the habitable zone. During episodes of high eccentricity, the time-averaged distance *d* of planet *b* from its host star, $d = a\,(1 + e^2/2)$, remained in the habitable zone. A higher eccentricity increases a planet's time-averaged distance, simply because a planet's orbital velocity decreases with orbital distance. A higher eccentricity does, of course, cause greater extremes. If, at perihelion, planet *b* was sufficiently heated so that it lost some of its water, then over time its water content could have evaporated. However, spending a larger time-averaged fraction of its orbit far from the star might have counteracted the increased heating at perihelion. During times of high eccentricity, planet *c*'s time-averaged orbital radius was increased to 1.41 AU, just beyond the habitable zone. Over the course of an orbit, water on its surface may alternately have frozen and thawed. Williams and Pollard (2002) found that the more important quantity for climate stability is the mean stellar flux received on the planetary surface, rather than the time spent in the habitable zone. Detailed models of orbit-climate interaction (e.g. Williams and Pollard 2002, 2003; Armstrong et al. 2004) are needed to assess the habitability of such planets.

---

[5]The inner and outer boundaries of the habitable zone are uncertain (e.g., Franck et al 2000, Mischna et al 2000). Thus, our choice of 0.9-1.4 AU is somewhat arbitrary.

## 5. Conclusions

We have investigated water delivery and planetary habitability in five simulations—first described in Raymond et al. (2006a)—with five to ten times more particles than in previous simulations. The planets that formed in these simulations were qualitatively similar to those formed in previous simulations. Following the results of Morbidelli et al. (2000), Lunine et al. (2003), and Raymond et al. (2004), we assumed an initial water gradient in the protoplanetary disk such that bodies originating at 1 AU were dry, while those past 2.5 AU contained 5% water. We discovered that *every* planet we formed was delivered at least five Earth oceans of water (1 ocean = $1.5 \times 10^{24}$ g is the amount of water on the Earth's surface). We formed several planets with a large amount of water, reminiscent of the ``water world'' planets formed in previous papers, notably Raymond et al. (2004, 2006c).

We propose a bimodal model for water delivery to Earth-like planets, which is a direct outgrowth of our simulations that included a larger number of particles than the pioneering simulations of Morbidelli et al. (2000). We suggest that terrestrial planets accrete a comparable amount of water in the form of a few water-rich Moon- to Mars-size planetary embryos and millions of km-size planetesimals. For this reason, the amount of water brought in by the large embryos in our simulation was smaller than in Morbidelli et al. (2000), and hence, the fraction of water supplied by small bodies was much larger. However, the overall behavior of our simulations—their sensitivity to initial conditions and, hence, variety of possible results—echoes the results of Morbidelli et al. (2000), even if we find somewhat smaller sensitivities by virtue of having a larger number of particles.

The water content of planets will vary because the number of embryos is relatively small, and hence, their water contribution is heavily influenced by initial conditions. In this respect our results are consistent with Morbidelli et al. (2000). However, water delivery from smaller planetesimals is statistically robust and should supply terrestrial planets with a significant water source of perhaps 3-10 oceans. We also expect planetary water content to vary systematically with orbital distance (as shown in Raymond et al., 2004) and with planet mass – future simulations will quantify the dependence on these parameters.

Our model applies to relatively dynamically calm environments such as cases with gas giant planets on circular orbits, or simply lower-mass giant planets. The prevalence of such conditions in the galaxy is not known, though recent observations have found systems containing only Neptune-mass planets (e.g., Lovis et al., 2006). In addition, the population of debris disks is anti-correlated with the known giant planets (Greaves et al., 2006), which suggests that terrestrial planets may often form in the absence of giant planets. Indeed, such dynamically calm conditions may be common in the Galaxy.

Our model argues that each of the Solar System terrestrial planets accreted a significant amount of water. Thus, the differences between planets' current water contents are likely due to water depletion, via processes such as loss during large impacts (e.g., Genda and Abe 2005) and hydrodynamical escape at early times (Matsui and Abe 1986).

Several factors not addressed here are relevant for water delivery. These include properties of the protoplanetary disk such as the disk's mass distribution (see Chambers and Cassen 2002, Raymond et al., 2005b), the total disk mass (see Raymond et al., 2006b), the location of the snow line in the disk (see, e.g., Lecar et al., 2006), and the location and orbits of giant planets (see Chambers 2003, Raymond et al., 2004, Raymond 2006). In addition, we have not considered alternate water sources such as comets, which were examined in detail by Morbidelli et al. (2000).

**Acknowledgments**
We thank Francis Nimmo, Hal Levison, and two anonymous referees for useful comments. We are grateful to NASA Astrobiology Institute for funding, through the University of Washington, Virtual Planetary Laboratory and University of Colorado lead teams. S.R. was partially supported by an appointment to the NASA Postdoctoral Program at the University of Colorado Astrobiology Center, administered by Oak Ridge Associated Universities through a contract with NASA. J.L. and T.Q. acknowledge the support of the International Space Science Institute (ISSI), Bern, Switzerland. Some of these simulations were run under CONDOR.[6]

---

[6]Condor is publicly available at www.cs.wisc.edu/condor

Table 1 -- Properties of **(potentially habitable)** planets formed[1]

| Simulation | Planet | a (AU) | $\bar{e}$[2] | $\bar{\iota}$ (°)[3] | $M(M_\oplus)$ | W.M.F. | Oceans[4] | Radius (km) | FeM.F. |
|---|---|---|---|---|---|---|---|---|---|
| 0 | a | 0.55 | 0.05 | 2.8 | 1.54 | $2.6 \times 10^{-3}$ | 15 | 7170 | 0.32 |
|  | **b** | **0.98** | **0.04** | **2.4** | **2.04** | **$8.4 \times 10^{-3}$** | **66** | **7730** | **0.28** |
|  | c | 1.93 | 0.06 | 4.6 | 0.95 | $9.1 \times 10^{-3}$ | 33 | 6290 | 0.28 |
| 1a | a | 0.58 | 0.05 | 2.7 | 0.93 | $8.3 \times 10^{-3}$ | 30 | 6250 | 0.31 |
|  | **b** | **1.09** | **0.07** | **4.1** | **0.78** | **$5.5 \times 10^{-3}$** | **17** | **5960** | **0.30** |
|  | c | 1.54 | 0.04 | 2.6 | 1.62 | $1.2 \times 10^{-2}$ | 75 | 7270 | 0.26 |
| 1b | a | 0.52 | 0.06 | 8.9 | 0.60 | $7.2 \times 10^{-3}$ | 17 | 5560 | 0.31 |
|  | **b** | **1.12** | **0.05** | **3.5** | **2.29** | **$6.7 \times 10^{-3}$** | **60** | **7980** | **0.29** |
|  | c | 1.95 | 0.09 | 9.7 | 0.41 | $3.8 \times 10^{-3}$ | 6 | 5010 | 0.28 |
| 2a | a | 0.55 | 0.08 | 2.6 | 1.31 | $9.3 \times 10^{-4}$ | 5 | 6860 | 0.33 |
|  | **b** | **0.94** | **0.13** | **3.4** | **0.87** | **$8.6 \times 10^{-3}$** | **29** | **6140** | **0.29** |
|  | **c** | **1.39** | **0.11** | **2.4** | **1.46** | **$6 \times 10^{-3}$** | **34** | **7060** | **0.29** |
|  | d | 2.19 | 0.08 | 8.8 | 1.08 | $1.8 \times 10^{-2}$ | 75 | 6510 | 0.24 |
| 2b | a | 0.61 | 0.18 | 13.1 | 2.60 | $7.1 \times 10^{-3}$ | 71 | 8260 | 0.30 |
|  | b | 1.72 | 0.17 | 0.5 | 1.63 | $2 \times 10^{-2}$ | 126 | 7280 | 0.22 |
|  |  |  |  |  |  |  |  |  |  |
| Mercury[5] |  | 0.39 | 0.19 | 7.0 | 0.06 | $1 \times 10^{-5}$ | 0 | 2436 | 0.68 |
| Venus |  | 0.72 | 0.03 | 3.4 | 0.82 | $5 \times 10^{-4}$ | 1.5 | 6052 | 0.33 |
| **Earth** |  | **1.0** | **0.03** | **0.0** | **1.0** | **$1 \times 10^{-3}$** | **4** | **6378** | **0.34** |
| Mars |  | 1.52 | 0.08 | 1.9 | 0.11 | $2 \times 10^{-4}$ | 0.1 | 3400 | 0.29 |

1. Planets are defined to be >0.2 Earth masses. Shown in bold are bodies in the habitable zone, defined to be between 0.9 and 1.4 AU. This is slightly wider than the most conservative habitable zone of Kasting et al. (1993).
2. Mean eccentricity averaged during the last 50 Myr of the simulation.
3. Mean inclination averaged during the last 50 Myr of the simulation.
4. Amount of planetary water in units of Earth oceans, where 1 ocean = $1.5 \times 10^{24}$, ($\approx 2.6 \times 10^{-4}$ $M_\oplus$) is the amount of water on Earth's surface. Earth's mantle contains about 1-10 oceans of water (see table 1 from Lecuyer et al., 1998).
5. Orbital values for the Solar system planets are 3 Myr averaged values from Quinn et al. (1991). Water contents are from Morbidelli et al. (2000). Iron values are from Lodders and Fegley (1998). The Earth's water content lies between 1 and 10 oceans -- here we assume a value of 4 oceans. See text for discussion.

Table 2 – Water Delivery to **(potentially habitable)** planets[1]

| Sim.-*planet* | a (AU) | M ($M_\oplus$) | N (oceans) | Water-delivering impacts (total)[2] | Water-bearing Embryos(plan'ls)[3] | % $H_2O$ from embryos[4] | Plan'l Oceans[5] | Mass ($M_\oplus$) from >2.5 AU accreted after last giant impact (% of total mass) |
|---|---|---|---|---|---|---|---|---|
| 0-*a* | 0.55 | 1.54 | 15 | 14 (18) | 1 (13) | 0.07 | 14 | 0.08 (5%) |
| **0-*b*** | **0.98** | **2.04** | **66** | **24 (60)** | **7 (17)** | **0.60** | **26** | **0.03 (1.6%)** |
| 0-*c* | 1.93 | 0.95 | 33 | 12(39) | 3(9) | 0.69 | 10 | 0.04 (4%) |
| 1a-*a* | 0.58 | 0.93 | 30 | 14 (25) | 2 (12) | 0.48 | 16 | 0.02 (2.7%) |
| **1a-*b*** | **1.09** | **0.78** | **17** | **7 (14)** | **1 (6)** | **0.43** | **10** | **0.09 (11%)** |
| 1a-*c* | 1.54 | 1.62 | 75 | 22 (64) | 4 (18) | 0.54 | 35 | 0.01 (0.4%) |
| 1b-*a* | 0.52 | 0.60 | 17 | 10 (14) | 1 (9) | 0.35 | 11 | 0.02 (3%) |
| **1b-*b*** | **1.12** | **2.29** | **60** | **24 (52)** | **7 (17)** | **0.66** | **20** | **0.02 (1%)** |
| 1b-*c* | 1.95 | 0.41 | 6 | 5 (5) | 1 (4) | 0.20 | 5 | 0.01 (3%) |
| 2a-*a* | 0.55 | 1.31 | 5 | 13 (16) | 1(12) | 0.12 | 4 | 0.02 (1%) |
| **2a-*b*** | **0.94** | **0.87** | **29** | **8 (10)** | **1(7)** | **0.94** | **2** | **0.01 (0.7%)** |
| **2a-*c*** | **1.39** | **1.46** | **34** | **11 (13)** | **3 (8)** | **0.98** | **1** | **0.003 (0.2%)** |
| 2a-*d* | 2.19 | 1.08 | 75 | 9 (12) | 4 (5) | 0.96 | 3 | -- |
| 2b-*a* | 0.61 | 2.60 | 71 | 9 (24) | 2 (7) | 0.96 | 3 | -- |
| 2b-*c* | 1.72 | 1.63 | 126 | 11 (24) | 3 (8) | 0.98 | 3 | -- |
| | | | | | | | | |
| **Earth**[6] | **1.0** | **1.0** | **1-10** | **?** | **?** | **?** | **?** | **<1-2%** |

1. Shown in bold are bodies in the habitable zone, between 0.9 and 1.4 AU.
2. Number of water-delivering impacts. Shown in parentheses is the total number of bodies incorporated in these impacts from past 2 AU.
3. Number of water-bearing planetary embryos, defined to be at least 0.05 $M_\oplus$. Shown in parentheses is the number of water-bearing planetesimals.
4. Fraction of the total water content delivered in the form of large bodies (embryos).
5. Number of oceans of water delivered to the planet in the form of planetesimals.
6. The amount of water in Earth's mantle is highly uncertain – see, e.g., Lecuyer et al. (1998). Siderophile abundances in the mantle imply that at most 1-2 percent of carbonaceous material impacted the Earth after its last core-forming event (Drake and Righter 2002).

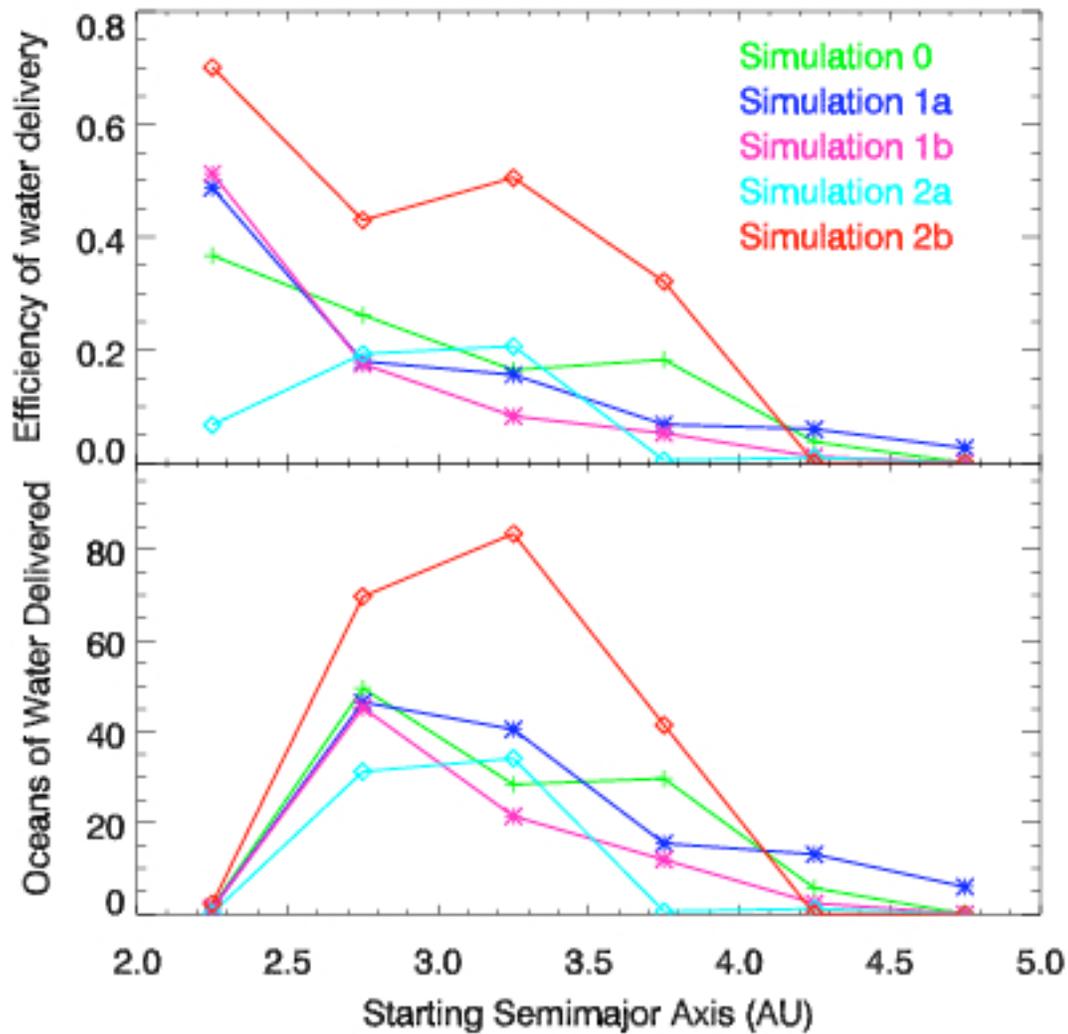

Figure 1 -- <u>Top</u>: Efficiency of water delivery in each of our five simulations. Each point represents the fraction of the (water-rich) material in a given 0.5 AU-wide bin to have been delivered to the surviving terrestrial planets inside 2 AU. <u>Bottom</u>: Source of water for the terrestrial planets in each simulation. Each point shows the number of oceans of water delivered from a given 0.5 AU-wide region to the surviving terrestrial planets inside 2 AU. Planet *d* from simulation 2a is not included in either panel.

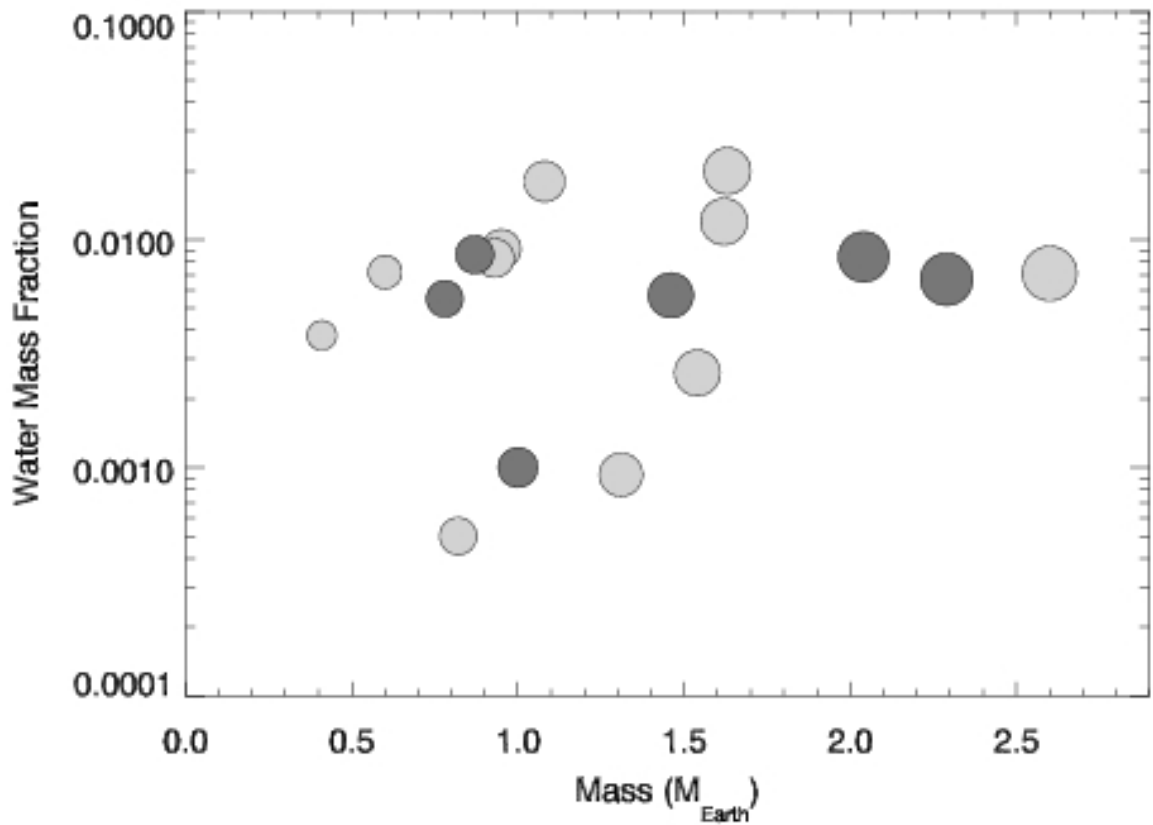

Figure 2 -- Water mass fraction of the surviving terrestrial planets as a function of their final mass. The size of each body is proportional to its relative physical size, and the darker planets are those located in the habitable zone.

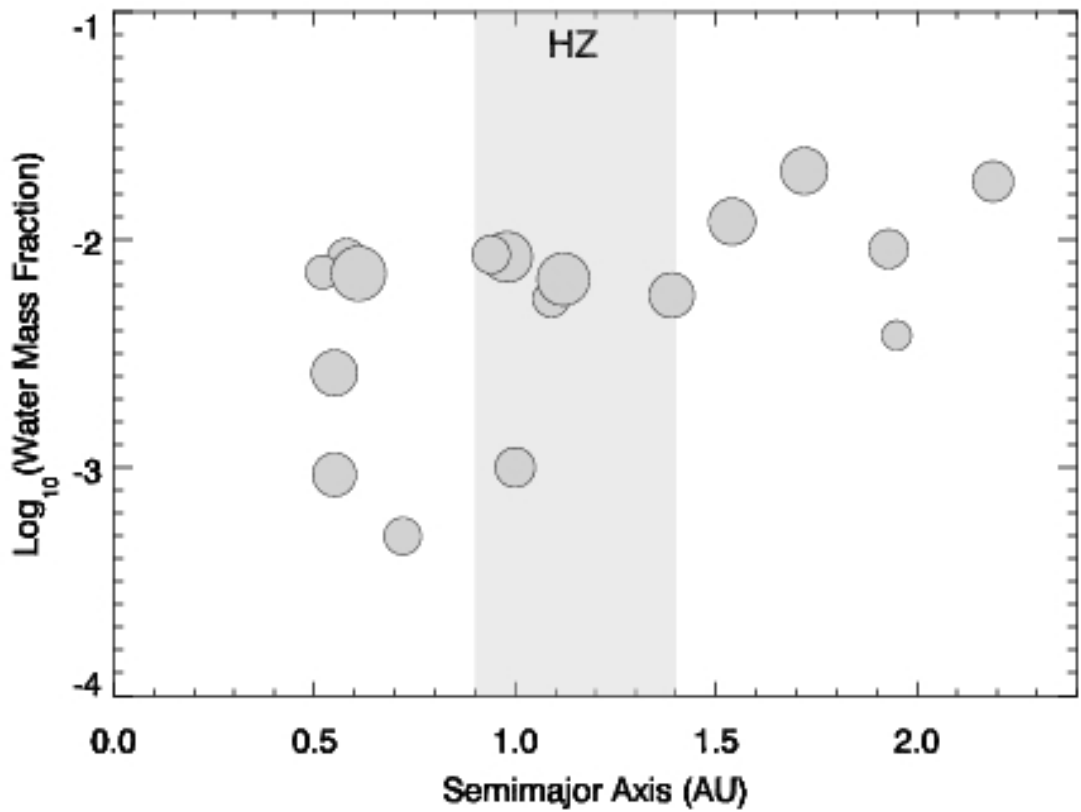

Figure 3 -- Water mass fraction of the surviving terrestrial planets as a function of their final orbital semimajor axis. The size of each body is proportional to its relative physical size. The habitable zone is shaded.

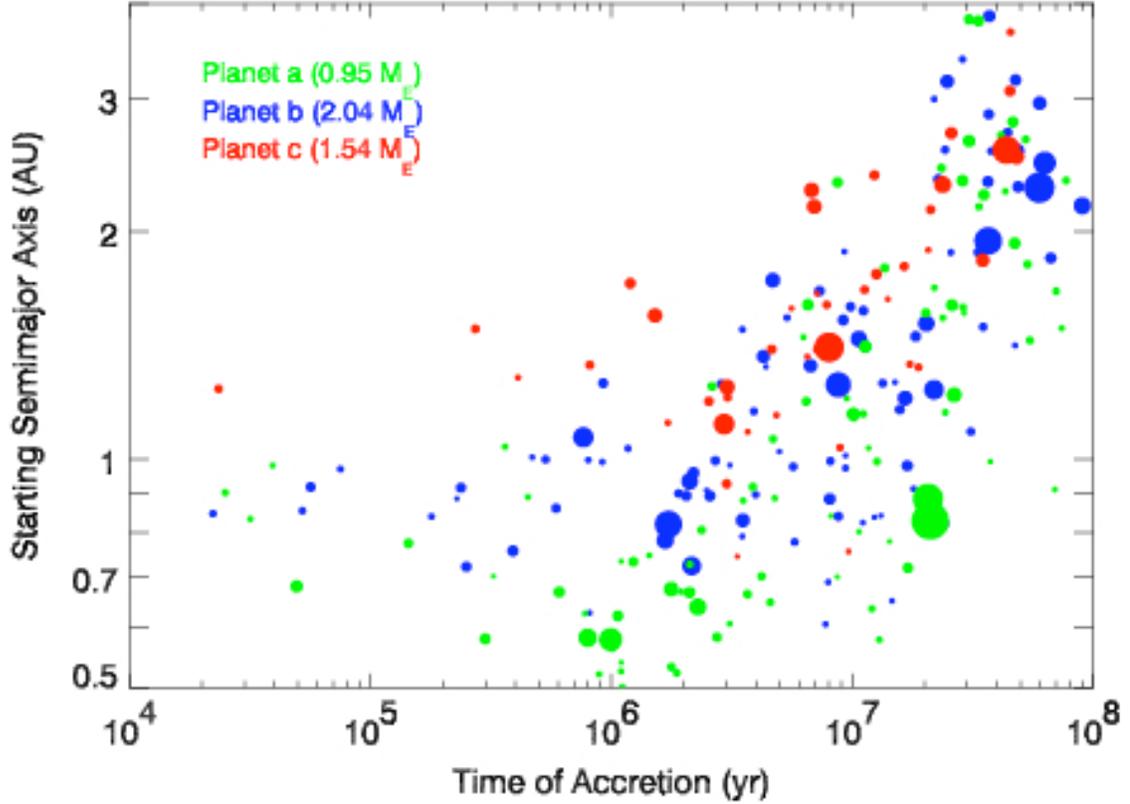

Figure 4 -- The timing of accretion of bodies from different initial locations for simulation 0. All bodies depicted in green were accreted by planet *a*, all blue bodies were accreted by planet *b*, and all red bodies were accreted by planet *c*. The relative size of each circle indicates its actual relative size. Impactors which had accreted other bodies are given their mass-weighted starting positions. Reproduced from Raymond et al. (2006a).

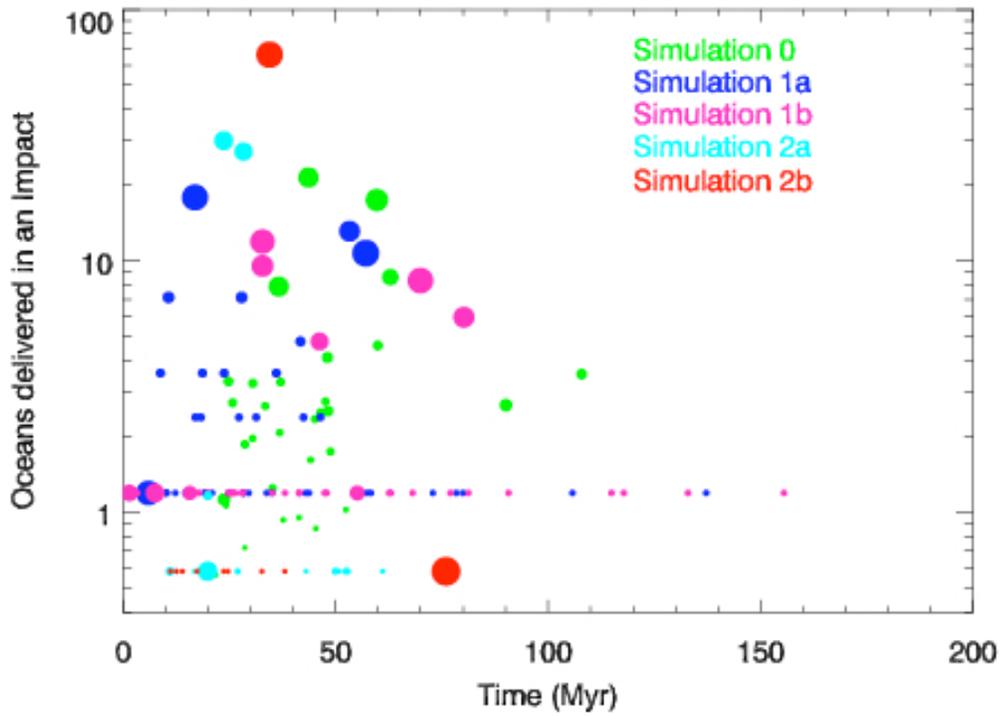

Figure 5 -- The amount of water (in units of Earth oceans, where 1 ocean is $1.5 \times 10^{24}$ grams) delivered to the terrestrial planets in each impact, shown through time. The size of each point represents the relative physical size of the impactor. The water contained in a planetesimal in simulations 1a and 1b is evident.

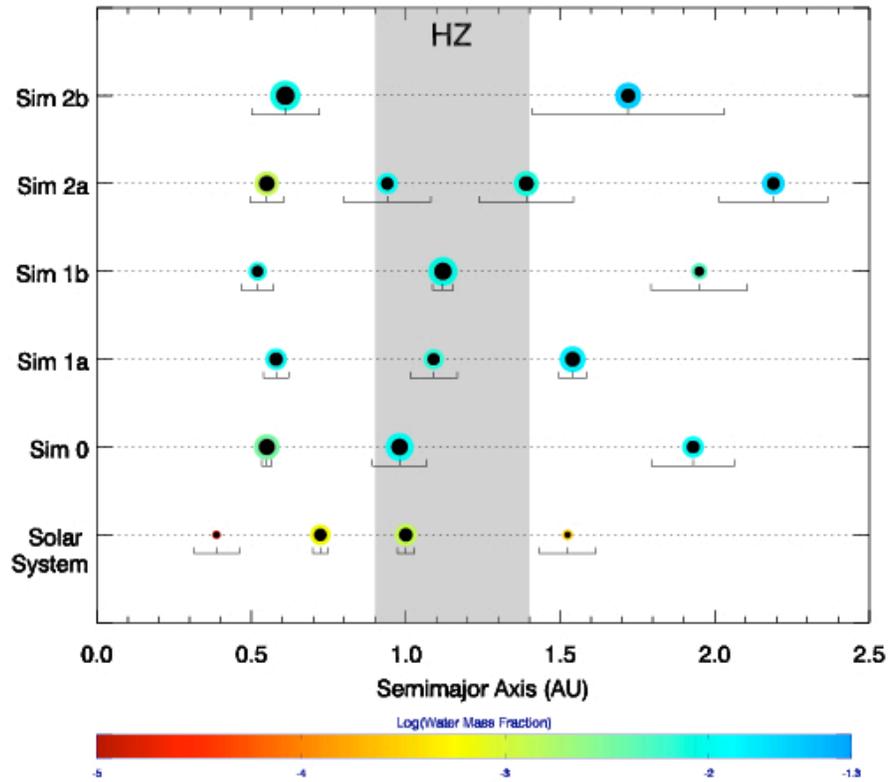

Figure 6 -- Final configurations of all five high-resolution simulations, with the Solar system terrestrial planets shown for scale. The relative size of each body corresponds to its relative physical size, and the color of each body represents its water mass fraction. The size of the dark inner region corresponds to the relative size of each planet's iron core. The shaded region (labeled "HZ") represents the habitable zone.

Figure 7 -- Eccentricities of planets *b* and *c* from simulation 2a as a function of time during a two

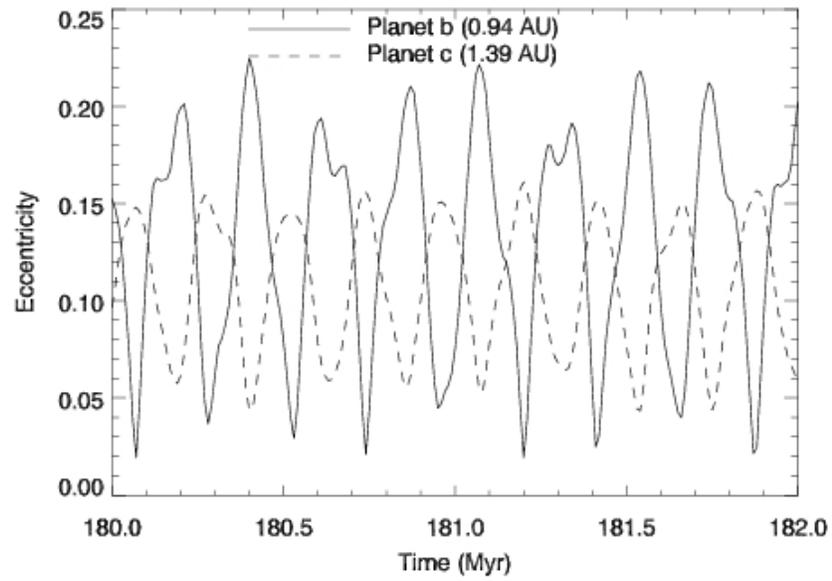

million year interval late in the simulation.

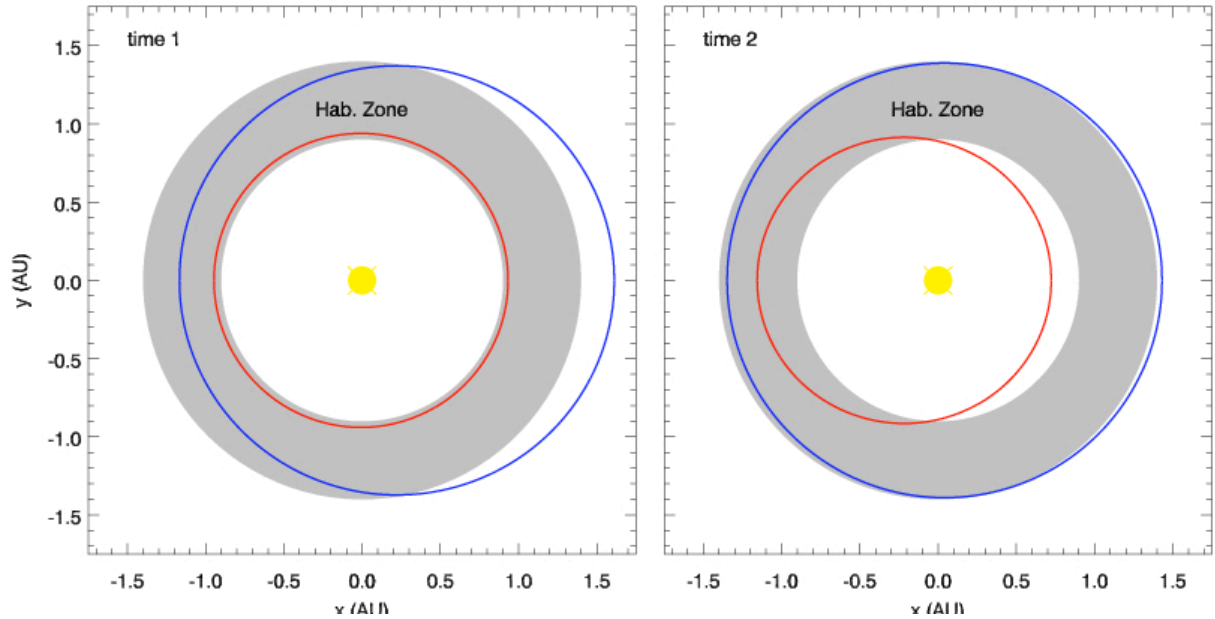

Figure 8 -- The orbits of planets *b* (red) and *c* (blue) from simulation 2a at two different times. During time 1 (left panel), planet *b*'s eccentricity is low and planet *c*'s is high. During time 2 (right panel), roughly 100 kyr after time 1 (see Fig. 7), planet *b*'s eccentricity is high and planet *c*'s is low. The habitable zone between 0.90 and 1.40 AU is shaded. The Sun is in yellow at the origin (not to scale).